\documentclass[preprint,prd,showpacs,amsmath,amssymb,amsthm,nofootinbib]{revtex4}
\usepackage{amssymb}
\usepackage{latexsym}
\usepackage{epsfig}
\usepackage{subfigure}
\usepackage{makecell}
\usepackage{amsmath}
\usepackage[colorlinks=true,linkcolor=red]{hyperref}
\newcommand{\bea}{\begin{eqnarray}}
\newcommand{\eea}{\end{eqnarray}}
\newcommand{\beq}{\begin{equation}}
\newcommand{\eeq}{\end{equation}}

\begin{document}

\title{Analytical approximations for primordial power spectra in a spatially closed emergent universe}
\author{Qihong Huang$^{1}$\footnote{Corresponding author: huangqihongzynu@163.com}, Kaituo Zhang$^{2}$, Zhenxing Fang$^{1}$ and Feiquan Tu$^{1}$}
\affiliation{
$^1$ School of Physics and Electronic Science, Zunyi Normal University, Zunyi 563006, China\\
$^2$ Department of Physics, Anhui Normal University, Wuhu, Anhui 241000, China
}

\begin{abstract}
The emergent universe scenario was proposed to solve the big bang singularity by suggesting that the universe originates from an Einstein static state and then evolves into a subsequently inflationary era. Thus, to find the relic of the existence of the Einstein static state becomes a crucial work. In this paper, we derive analytical approximation of the primordial power spectra and analyze the CMB TT-spectra for the spatially closed emergent universe. After analyzing the CMB TT-spectrum of the emergent universe scenario, we find that both the CMB TT-spectra produced by the Einstein static state followed by the ultraslow-roll inflationary epoch (method I) and by a special evolution of the scale factor in the emergent scenario as $a=a_{0}+A e^{H_{0}t}$(method II) are suppressed at $l<30$, and their spectra are nearly identical. Additionally, by comparing the spectra of the emergent universe scenario with the ones of the ultraslow-roll inflationary model in the closed universe, we find that the CMB TT-spectrum of the emergent universe is similar to the one of the inflationary model with the special case $\eta_t = \eta_{max}$.
\end{abstract}

\maketitle

\section{Introduction}

Inflation~\cite{Guth1981, Linde1982, Albrecht1982}, invoked to resolve several important issues encountered in the standard cosmology, has been a dominant paradigm of the universe in current cosmology, and it can provide important clues for large scale structure formation of the universe. However, the big bang singularity problem still exists. To solve this problem, Ellis \textit{et al.}~\cite{Ellis2004a, Ellis2004b} proposed a model named emergent universe. In this scenario, the universe initially stays in a past-eternal Einstein static state and then evolves into a subsequently inflationary epoch eventually. Thus, the big bang singularity can be avoided. Subsequently, the emergent universe scenario has been studied in various modified theories~\cite{Carneiro2009, Huang2014, Campo2007, Campo2009, Wu2011, Li2013, Mulryne2005, Lidsey2004, Parisi2007, Canonico2010, Wu2009, Bag2014, HuangQ2015, Bohmer2004, Atazadeh2014, Zhang2016, Zhang2010, Zhang2012, Lidsey2006, Gruppuso2004, Gergely2002, Atazadeh2014a, Zhang2014, Heydarzade2016, Heydarzade2015, HuangH2015, Bohmer2009, Wu2010, Bohmer2010, Khodadi2016, Bohmer2015, Atazadeh2015, Atazadeh2017, Bohmer2013, Tawfik2016, Khodadi2015, Odrzywolek2009, Clifton2005, Vilenkin2013, Aguirre2013, Mithani2012, Cai2014, Cai2012, Mousavi2017, Miao2016, Seahra2009, Li2017, Shabani2017, Huang2018, HuangQ2018, Shabani2019, Sharif2019, Li2019, Huang2020, Mukherjee2006, Beesham2009, Paul2010, Debnath2011, HuangQ2018a, HuangQ2018b, Alesci2017, Alesci2018}.

Observations of the cosmic microwave background (CMB) radiation suggest that the suppression of CMB TT-spectrum exists at large scales (multipoles $l<40$), which was first observed by COBE~\cite{Smoot1992}, then by WMAP~\cite{Spergel2007}, and recently confirmed by Planck 2018~\cite{Planck2020}. This observational result is intriguing since the low multipoles $l$ modes in CMB TT-spectrum at present time correspond to very large wavelength modes. These large wavelength modes have not been affected by the late-time evolution of the universe due to their superhorizon sizes from inflation period to the present. So, the power suppression observed in the CMB TT-spectrum at low multipoles $l$ can correspond to the physics of the very early universe.

To explain the power suppression of CMB TT-spectrum, several approaches are proposed. One approach is to introduce a cutoff in the primordial power spectrum~\cite{Bridle2003, Contaldi2003, Cline2003, Liu2013, Labrana2015}. This cutoff can be related to string physics~\cite{Gil2003, Dudas2012}, the bouncing universe~\cite{Liu2013, Piao2004}, or a fast roll stage of the inflaton field~\cite{Contaldi2003}. The most prominent feature of the primordial power spectrum is a sharp and infrared cutoff on the horizon scale. Another approach is to consider the inflation in a curved universe~\cite{Bonga2016, Handley2019, Thavanesan2021}, which predicts a slightly red-tilted power spectrum of the primordial scalar perturbation and is consistent with the recent cosmological observations. Although the effects of curvature on the CMB TT-spectrum is limited to low multipoles $l$, it is sensitive to the initial conditions of inflation. In addition, some models were also proposed to explain the power suppression of CMB TT-spectrum, such as, pre-inflation~\cite{Dudas2012, Cai2015}, pre-inflationary bounce~\cite{Cai2018}, anisotropic universe~\cite{Campanelli2006, Campanelli2007, Campanelli2009}, non-flat XCDM inflation model~\cite{Ooba2018}, emergent universe scenario~\cite{Labrana2015},and so on.

Recently, the primordial power spectra of the emergent universe were studied in general relativity~\cite{Labrana2015, Martineau2018} and quantum reduced loop gravity~\cite{Olmedo2018}. And the emergent universe scenario was used to explain the power suppression of CMB TT-spectrum by assuming the Einstein static state as a superinflating phase, which exists before the onset of inflation~\cite{Labrana2015}. In this work, based on general relativity, the emergent universe scenario in the closed FLRW spacetime was studied. The results showed that the superinflationary epoch of emergent universe scenario could produce a power suppression of CMB TT-spectrum at large scale. In this case, the spatial curvature is only responsible for the superinflationary epoch, i.e. it only ensure the existence of Einstein static universe, not the primordial perturbations. Although the study using a statistically powerful Planck likehood suggests that the Planck temperature and polarization spectra are consistent with a spatially flat Universe~\cite{Efstathiou2020}, it is noteworthy that the spatial curvature plays an important role in cosmology. The spatial curvature can suppress the primordial power spectrum and the CMB TT-spectrum in the inflationary models~\cite{Bonga2016, Handley2019, Thavanesan2021}. And the positive spatial curvature is favored by Planck 2018 data without the lensing likelihood~\cite{Planck2020, Handley2021}, full-shape Galaxy power spectra~\cite{Glanville2022}, Planck PR4~\cite{Rosenberg2022}, and so on. Similar conclusions are also reached by using different statistical techniques~\cite{Handley2019} and different arguments~\cite{Ooba2018a, Park2019}.  When the positive curvature is introduced, the enhanced lensing amplitude in CMB power spectra can be explained naturally~\cite{Valentino2020}. Thus, in this work, we plan to explore whether the positive spatial curvature in emergent universe is helpful to explain the power suppression of the CMB TT-spectrum.

It is notable that, in order to get an analytical approximation of primordial power spectrum in curved universe, the inflationary model was approximated as a kinetically dominated epoch followed by a ultraslow-roll epoch~\cite{Thavanesan2021}. And the significant suppression of the CMB TT-spectrum was found in this inflationary model with case of $\eta_t = 0.01\eta_{max}$ and $\eta_t = 0.03\eta_{max}$ in the closed spacetime. However, it is unclear whether such an suppression can exist in the spatially closed emergent universe.

The paper is organized as follows. In Section II, we briefly review the background equations and Mukhanov-Sasaki equation in a curved spacetime. In Section III, we solve the curved Mukhanov-Sasaki equation and obtain the analytical primordial power spectra of the spatially closed emergent universe. In Section IV, we plot the CMB TT-spectra of the emergent universe scenario in a closed universe. Finally, our main conclusions are drawed in Section V.

\section{Background}

In this section, we begin with the action for a single-component scalar field minimally coupled to a curved spacetime
\beq\label{S}
S=\int d^4 x \sqrt{\left| g \right|}\Big[\frac{1}{2}R+\frac{1}{2}\nabla^{\mu}\phi\nabla_{\mu}\phi-V(\phi)\Big]
\eeq
where $R$ is the Ricci curvature scalar, $\phi$ is the scalar field, and $V(\phi)$ is the potential.

Since the emergent universe scenario generally requires a positive curvature, we consider a closed universe $(K=1)$ in the following. For the closed FLRW universe, the perturbed metric in the Newtonian gauge is
\bea\label{ds}
&& ds^2=a(\eta)^2 \big[(1+2\Phi)d\eta^{2}-(1-2\Psi)c_{ij}dx^i dx^j\big],\nonumber\\
&& c_{ij}dx^i dx^j=\frac{dr^2}{1-r^2}+r^2 (d\theta^2+sin^2 \theta d\varphi^2),
\eea
where $a$ denotes the scale factor, $\eta$ represents the conformal time, and the longitudinal metric perturbation $\Phi$ and curvature metric perturbation $\Psi$ characterize the scalar perturbations.

Varying the action~(\ref{S}) with respect to the metric tensor $g^{\mu\nu}$ and the scalar field $\phi$, we obtain the Einstein field equation and the scalar field equation. The $0-0$ components of the Einstein field equation and the scalar field equation can be expressed as
\bea\label{h01}
&& \mathcal{H}^2+1=\frac{\kappa^2}{3}\Big[\frac{1}{2}\phi'^{2}+a^2 V(\phi)\Big],
\eea
\bea\label{h02}
&& \phi''+2\mathcal{H}\phi'+a^2 \frac{d}{d\phi}V(\phi)=0,
\eea
where $\kappa^2=8\pi G$, $\mathcal{H}=a'/a$ indicates the conformal Hubble parameter, and $'$ represents the derivatives with respect to the conformal time defined by $d\eta=dt/a$. Combining Eqs.~(\ref{h01}) and ~(\ref{h02}), we get
\bea\label{eqh1}
\mathcal{H'}+2\mathcal{H}^2+2=\kappa^2a^2 V(\phi),
\eea
\bea\label{eqh2}
\mathcal{H'}-\mathcal{H}^2-1=-\frac{1}{2}\kappa^2\phi'^2.
\eea

In order to analyze the primordial power spectrum in emergent universe scenario, we first introduce a gauge-invariant comoving curvature perturbations $\mathcal{R}$ defined by
\bea
\mathcal{R}=\Psi+\frac{\mathcal{H}}{\phi'}\delta\phi,
\eea
which satisfies an equation of motion named Mukhanov-Sasaki equation. In the curved universe, the Mukhanov-Sasaki equation was generalized to~\cite{Handley2019, Thavanesan2021}
\bea
(\mathcal{D}^2-K \varepsilon)\mathcal{R}''+\Big[\Big(\frac{\phi'^2}{\mathcal{H}}+\frac{2\phi''}{\phi'}-\frac{2K}{\mathcal{H}}\Big)\mathcal{D}^2-2K\mathcal{H}\varepsilon\Big]\mathcal{R}'\nonumber\\
+\Big[-\mathcal{D}^4+K\Big(\frac{2K}{\mathcal{H}^2}-\varepsilon+1-\frac{2\phi''}{\phi'\mathcal{H}}\Big)\mathcal{D}^2+K^2 \varepsilon\Big]\mathcal{R}=0,
\eea
where
\bea
\mathcal{D}^2=\nabla_i \nabla^i+3K, \qquad \varepsilon=\frac{\phi'^2}{2\mathcal{H}^2}.
\eea
By defining two new variables
\bea
v=\mathcal{Z}\mathcal{R}, \qquad \mathcal{Z}=\frac{a \phi'}{\mathcal{H}}\sqrt{\frac{\mathcal{D}^2}{\mathcal{D}^2-K \varepsilon}},
\eea
and replacing the $\mathcal{D}^2$ operator with its associated scalar wave vector expression~\cite{Lesgourgues2014},
\bea
\mathcal{D}^2 \leftrightarrow -\mathcal{K}^2(k)+3K,
\eea
\bea
\mathcal{K}^2(k)=\Bigg\{
\begin{array}{lll}
k(k+2),\quad & k=2,3,..., \quad & K=+1\\
k^2,\quad & k>0, \quad & K=0,-1
\end{array}
\eea
the curved Mukhanov-Sasaki equation in momentum space can be expressed as
\bea\label{vvk}
v''_k+\Big[\mathcal{K}^2-\Big(\frac{\mathcal{Z}''}{\mathcal{Z}}+2K+\frac{2K \mathcal{Z}'}{\mathcal{H}\mathcal{Z}}\Big)\Big]v_k=0
\eea
For the closed universe, $K=1$ is required. After solving Eq.~(\ref{vvk}), we obtain the solution for the Mukhanov variables $v_k$. Then, we can derive the curved primordial power spectrum of the comving curvature perturbation $\mathcal{R}$
\bea\label{PR0}
\mathcal{P}_{\mathcal{R}}=\frac{k^3}{2\pi^2}\left| \mathcal{R}_k \right|^2 =\frac{k^3}{2\pi^2}\left| \frac{v_k}{\mathcal{Z}_k} \right|^2.
\eea

\section{Analytical primordial power spectra}

In emergent universe scenario, the universe first stems from a past-eternal classical Einstein static state universe, then exits naturally from this static state, and eventually evolves into an inflationary epoch~\cite{Ellis2004a, Ellis2004b}. It is assumed that the scalar field $\phi$ rolls on an asymptotically flat scalar potential $V(\phi)$ with a constant velocity $\phi'$, which provides the static conditions for the Einstein static universe. With the increasing of time, the scalar field exceeds a critical point, the scalar potential decreases slowly, and the universe departs from the Einstein static state and eventually enters into the inflationary epoch. In this paper, we use two different methods to discuss the evolution of scale factor $a$ in emergent universe scenario:

(1)Method I: we assume the Einstein static state region defined by $a'=0$, and then invoke an instantaneous transition to the ultraslow-roll inflationary epoch $\phi'^2<<a^2 V(\phi)$. To realize the instantaneous transition from the Einstein static state to the inflationary stage, some approaches had been proposed, for example, breaking the stability conditions~\cite{Wu2010, HuangQ2015}. In order to break the stability conditions, it often requires that the scalar potential or the equation of state varies with time very slowly. Once the time increases to the critical point, the stability conditions are broken. Then the universe exits from the Einstein static state and evolves into a subsequently inflationary epoch. This method was widely used in the emergent scenario~\cite{Campo2007, Wu2010, Zhang2014, HuangQ2015, Shabani2019, Huang2020}.

(2)Method II: we consider the evolution of the scale factor in the emergent scenario as ~\cite{Ellis2004a, Ellis2004b}
\bea
a(t)=a_{0}+A e^{H_{0} t},
\eea
which can be rewritten in the conformal time as~\cite{Labrana2015}
\bea
a(\eta)=\frac{a_0}{1-e^{a_0 H_0 \eta}}, \qquad \eta<0.
\eea
Since $a(\eta)\rightarrow a_{0}$ when $\eta\rightarrow-\infty$, the universe is asymptotic to an Einstein static state in the infinite past. With the increase of $\eta$, the universe gradually departs from the Einstein static state and then enters into the inflationary epoch when $\eta\rightarrow 0$.

\subsection{Method I}

In the Einstein static state regime, we adopt the Einstein static conditions $a'=0$ and $\mathcal{H}=0$, then the background equation~(\ref{eqh1}) becomes
\bea\label{a00}
a(\eta)=a_{0}=\sqrt{\frac{2}{\kappa^2 V_0}}.
\eea
Here, $a_0$ and $V_0$ are constants and represent the corresponding values in the static state. Under the Einstein static state conditions, the variable $\mathcal{Z}$ reduces to
\bea\label{e1}
\mathcal{Z}=a_0 \sqrt{2(k^2+2k-3)},
\eea
which is independent of time. Then, substituting ~(\ref{e1}) into ~(\ref{vvk}), we can write the curved Mukhanov-Sasaki equation~(\ref{vvk}) in the Einstein static state regime as
\bea
v''_k+k_{-}^2 v_k=0, \qquad k_{-}^2=k(k+2)-4,
\eea
with the solution
\beq
v_k(\eta)=A_{k} e^{i k_{-}\eta}+B_{k} e^{-i k_{-}\eta}.
\eeq
Using the normalization conditions of quantum mechanics $v_k v_k^{*'}-v_k^{'} v_k^{*}=i$, we obtain
\beq
B_{k}^{2}-A_{k}^{2}=\frac{1}{2k_{-}}.
\eeq

In order to fully determine the Mukhanov variables $v_{k}$, an initial condition for $v_{k}$ should be specified. This initial condition that to fully fix $v_{k}$ comes from vacuum selection. Recently, by choosing Bunch-Davies vacuum, Hamiltonian Diagonalisation, Renormalized Stress Energy Tensor, Right Handed Mode, and Frozen Initial Conditions as the quantum initial condition before inflation, the primordial power spectra and the CMB TT-spectra have been discussed in ~\cite{Gessey-Jones2021}. The results show that some choices of vacuum can distinguishable from others, but the Planck 2018 shows no significant evidence to favour any of the quantum vacuum. In the Einstein static state, the Hubble horizon is infinity since the scale factor $a$ is a constant and $H=0$. As a result, all observable perturbation modes are deeply inside the Hubble horizon $1/H$ and the comoving horizon $1/a H$. Thus, a standard choice is the Bunch-Davies vacuum~\cite{Martineau2018, Bonga2016, Baumann2009}, in which a positive frequency solution is obtained and the vacuum stays the minimum energy state. This choice requires the Mukhanov variables $v_{k}$ satisfies the constraint condition
\beq
v_k^{'2}+k_{-}^2v_k^2=0.
\eeq
Thus, we obtain the initial conditions
\beq
A_{k}=0, \qquad B_{k}=\sqrt{\frac{1}{2k_{-}}}
\eeq
and the solution
\bea\label{vk00}
v_k(\eta)=\sqrt{\frac{1}{2k_{-}}}e^{-i k_{-}\eta}.
\eea

\begin{figure*}[htp]
\begin{center}
\includegraphics[width=0.45\textwidth]{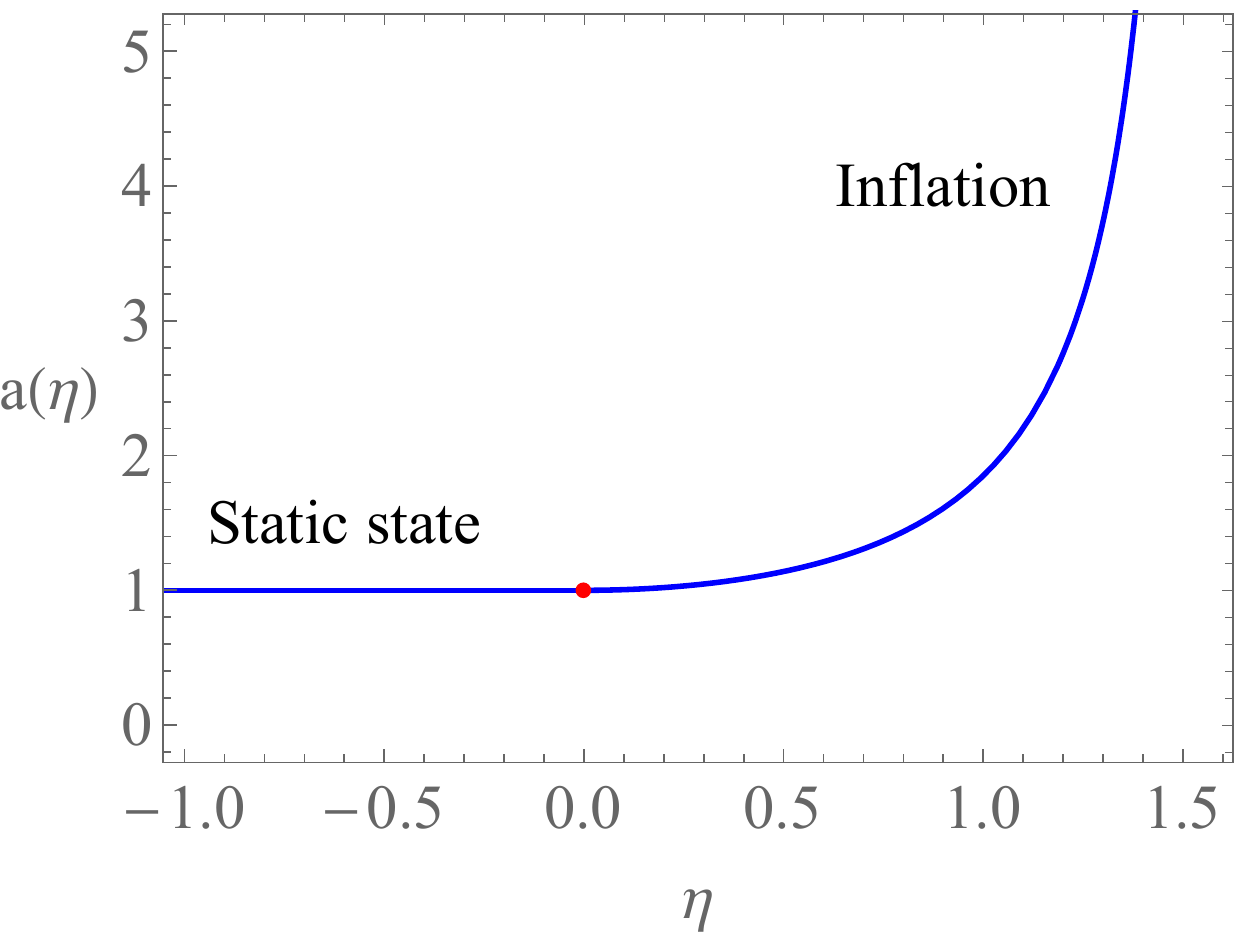}
\caption{\label{Fig01} Evolutionary curve of the scale factor $a$, where the scalar factor in Einstein static state has been set as $a_0=1$ and the transition time takes $\eta_t=0$. The red point represents the transition point.}
\end{center}
\end{figure*}

Then, following Ref.~\cite{Thavanesan2021}, for the ultraslow-roll regime $\phi'^2<<a^2 V(\phi)$, we consider the right hand side of Eq.~(\ref{eqh2}) to be zero and then Eq.~(\ref{eqh2}) becomes
\beq
\mathcal{H'}-\mathcal{H}^2-1=0,
\eeq
which gives
\bea\label{a001}
a(\eta)\sim\frac{1}{\cos(\eta)}.
\eea
Matching $a$ and $a'$ for these two solutions ~(\ref{a00}) and ~(\ref{a001}) at a transition time $\eta_t$, we get the evolutionary expression for the scale factor $a$ from the static state to the inflationary epoch
\bea\label{IIa}
a(\eta)=\Bigg\{
\begin{array}{ll}
a_0,\quad & \eta<\eta_t\\
\frac{a_0}{\cos(\eta-\eta_t)},\quad & \eta_t \leq \eta < \eta_t+\frac{\pi}{2}.
\end{array}
\eea
With $\eta$ approaching to $\eta_t+\frac{\pi}{2}$, the universe freezes out into the inflationary phase. The evolutionary curve of the scale factor $a$ is plotted in Fig.~(\ref{Fig01}).

In the ultraslow-roll regime, taking into consideration the constraint $\varepsilon=\frac{\phi'^2}{2\mathcal{H}^2}\approx 0$, we obtain~\cite{Thavanesan2021}
\bea
\frac{\mathcal{Z}''}{\mathcal{Z}}+2+\frac{2 \mathcal{Z}'}{\mathcal{H}\mathcal{Z}}\approx \frac{a''}{a}+3,
\eea
and the curved Mukhanov-Sasaki equation~(\ref{vvk}) takes the form
\bea
v''_k+\Big\{k_{+}^2-\frac{2}{\big[\eta-\big(\eta_t+\frac{\pi}{2}\big)\big]^2}\Big\}v_k=0, \qquad k_{+}^2=k(k+2)-\frac{8}{3}.
\eea
Solving these equation, we obtain the solution of the Mukhanov variable $v_k$
\bea\label{vkcd}
v_k(\eta)=\sqrt{\frac{\pi}{4}}\sqrt{\big(\eta_t+\frac{\pi}{2}\big)-\eta}\Big[C_k H^{(1)}_{3/2}\Big(k_{+}\big((\eta_t+\frac{\pi}{2})-\eta\big)\Big)+ D_k H^{(2)}_{3/2}\Big(k_{+}\big((\eta_t+\frac{\pi}{2})-\eta\big)\Big)\Big],\nonumber\\
\eea
where $H^{(1)}$ and $H^{(2)}$ are the Hankel functions of the first and second kinds.

To determine the coefficients of the Mukhanov variable $v_k$ in Eq.~(\ref{vkcd}), we match Eq.~(\ref{vk00}) and Eq.~(\ref{vkcd}) under the condition of the continuity of $v_k$ and $v_k^{'}$ at the transition time $\eta_t$ and then get
\bea
C_k=\frac{1}{4}e^{-i k_{-} \eta_t}\sqrt{\frac{1}{k_{-}}}\Big[i \pi k_{+} H^{(2)}_{1/2}\Big(\frac{\pi}{2}k_{+}\Big)+(-2i+\pi k_{-})H^{(2)}_{3/2}\Big(\frac{\pi}{2}k_{+}\Big)\Big],
\eea
\bea
D_k=-\frac{1}{4}e^{-i k_{-} \eta_t}\sqrt{\frac{1}{k_{-}}}\Big[i \pi k_{+} H^{(1)}_{1/2}\Big(\frac{\pi}{2}k_{+}\Big)+(-2i+\pi k_{-})H^{(1)}_{3/2}\Big(\frac{\pi}{2}k_{+}\Big)\Big].
\eea

\subsection{Method II}

In this method, we consider the evolution of the scale factor in the conformal time as follows~\cite{Labrana2015}
\bea\label{a01}
a(\eta)=\frac{a_0}{1-e^{a_0 H_0 \eta}}, \qquad \eta<0,
\eea
in which the scale factor $a$ is only asymptotically tends to a constant in the infinite past and the universe is not a truly static during the Einstein static region.

Following Ref.~\cite{Labrana2015} and using the expression of $a$, we can obtain
\bea\label{Z2Z}
\frac{\mathcal{Z}''}{\mathcal{Z}}+\frac{2\mathcal{Z}'}{\mathcal{H}\mathcal{Z}}\approx \frac{a''}{a}+1.
\eea
Combining Eqs.~(\ref{vvk}),~(\ref{a01}) and ~(\ref{Z2Z}), the curved Mukhanov-Sasaki equation~(\ref{vvk}) becomes
\bea
v''_k+\Big[k^2_0-(a_0 H_0)^2 e^{a_0 H_0 \eta}\frac{1+e^{a_0 H_0 \eta}}{(1-e^{a_0 H_0 \eta})^2}\Big]v_k=0, \qquad k_0^2=k(k+2)-3.
\eea
This equation can be solved by
\bea\label{vk0}
v_k(\eta)&&=\frac{1}{\sqrt{2k_0}}\frac{e^{-i k_0 \eta}}{1-e^{a_0 H_0 \eta}} \times \ _2F_1\Big(-1-\frac{i k_0}{a_0 H_0}-\sqrt{1-\big(\frac{k_0}{a_0 H_0}\big)^2},\nonumber\\
&&-1-\frac{i k_0}{a_0 H_0}+\sqrt{1-\big(\frac{k_0}{a_0 H_0}\big)^2}; 1-2\frac{i k_0}{a_0 H_0};e^{a_0 H_0 \eta}\Big),
\eea
where $_{2}F_{1}$ is the hypergeometric function, and the combination $a_0 H_0$ is a free parameter which can be set $a_0 H_0=0.0002Mpc^{-1}$~\cite{Labrana2015}. In the short wavelengths limit, the normalized positive frequency modes correspond to the minimal quantum fluctuations
\beq\label{vk0a}
v_k(\eta)\approx \frac{1}{\sqrt{2k_0}}e^{-i k_0 \eta},\quad a H \ll k_0.
\eeq
This result is also obtained in method I(i.e. Eq.~(\ref{vk00})). For $K=0$, we obtain $k^2_0=k^2$ and the solution of $v_k(\eta)$ reduces to the case in Ref.~\cite{Labrana2015}. It is interesting to note that the free parameter $a_0 H_0$ is avoided in method I.

\subsection{Primordial power spectra}

Substituting Eq.~(\ref{vkcd}) into Eq.~(\ref{PR0}), the primordial power spectrum of the comoving curvature perturbation $\mathcal{R}$ for method I is
\bea
\mathcal{P}_{\mathcal{R}}&&=\frac{k^3}{2\pi^2}\left| \mathcal{R}_k \right|^2 \approx \lim_{\eta\rightarrow\eta_t+\frac{\pi}{2}}\frac{1}{8 a^2 \pi^2 \varepsilon \big[\eta-\big(\eta_t-\frac{\pi}{2}\big)\big]^2}\frac{k^3}{k^3_{+}} \left| C_k-D_k \right|^2\nonumber\\
&&=A_s \frac{k^3}{k^3_{+}}\left| C_k-D_k \right|^2.
\eea
Here, similar to Ref.~\cite{Thavanesan2021}, the transition time parameter $\eta_t$, slow-roll parameter $\varepsilon$ and formally diverging parameters are absorbed into the scalar power spectrum amplitude $A_s$. In the short wavelengths limit, where $k_{+} \approx k_{-} \approx k$, we recover the standard scale-invariant spectrum
\bea
\left| C_k \right| \approx 1, \quad \left|D_k \right| \approx 0, \quad \mathcal{P}_{\mathcal{R}} \approx A_s.
\eea

Then, the analytical primordial power spectrum is parameterized as
\bea
\mathcal{P}_{\mathcal{R}}(k)=A_{s}\Big(\frac{k}{k_{*}}\Big)^{n_s-1}\frac{k^3}{k^3_{+}}\left| C_k-D_k \right|^2,
\eea
where $k_{*}=0.05 Mpc^{-1}$ corresponding to the pivot perturbation mode.

\begin{figure*}[htp]
\begin{center}
\includegraphics[width=0.45\textwidth]{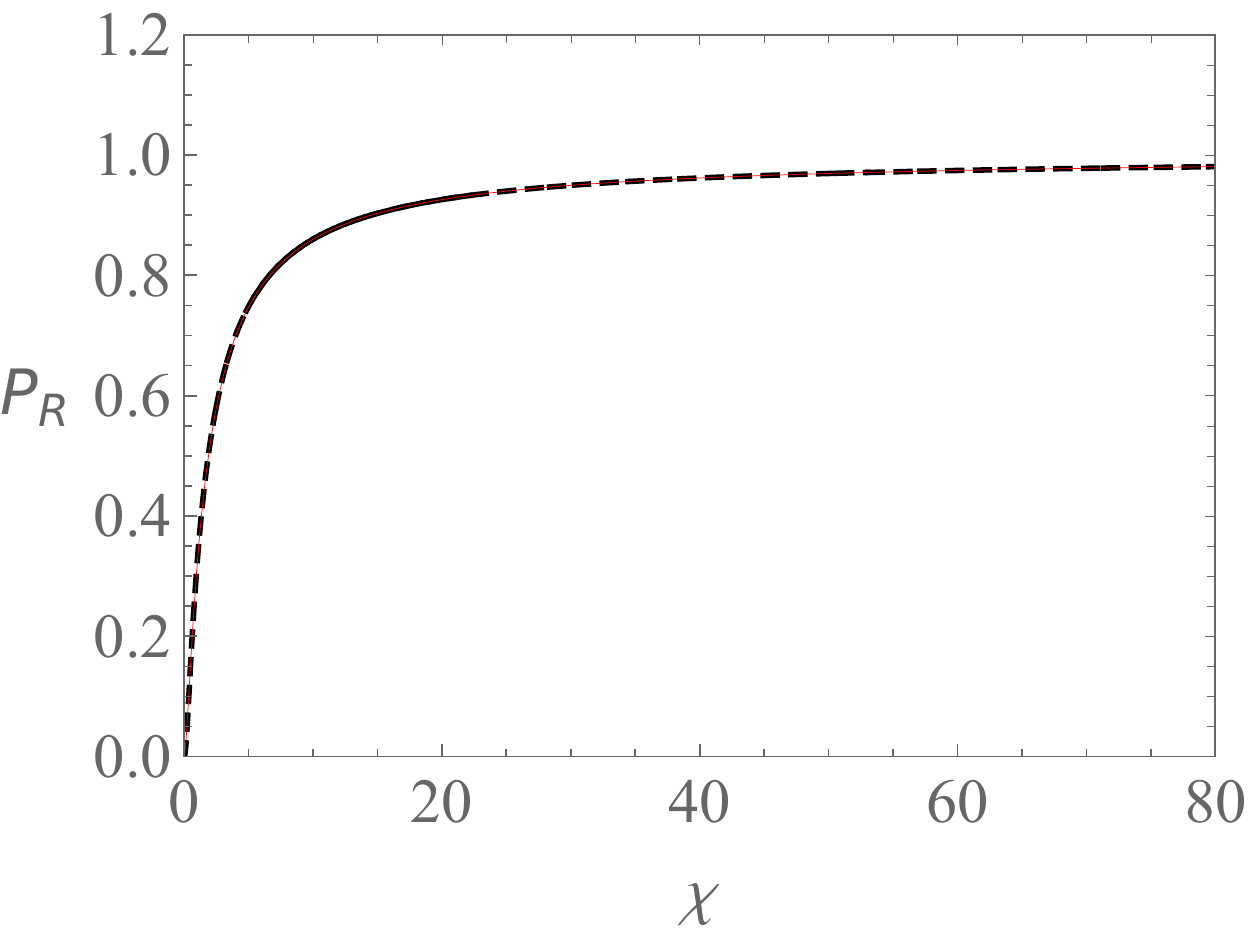}
\caption{\label{Fig02} Power spectra of $\mathcal{P}_{\mathcal{R}}$ for method II. The analytical spectrum Eq.~(\ref{P21}) is plotted by the black dashed line, while the approximate spectrum Eq.~(\ref{P22}) is depicted by the red one.}
\end{center}
\end{figure*}

For method II, after substituting Eq.~(\ref{vk0}) into Eq.~(\ref{PR0}), we obtain
\bea\label{P21}
\mathcal{P}_{\mathcal{R}} &&\sim 4\chi^2 \frac{\Gamma(1-2i\chi)}{\Gamma(2-i\chi-\sqrt{1-\chi^2})\Gamma(2-i\chi+\sqrt{1-\chi^2})}\nonumber\\
&&\times \frac{\Gamma(1+2i\chi)}{\Gamma(2+i\chi-\sqrt{1-\chi^2})\Gamma(2+i\chi+\sqrt{1-\chi^2})},
\eea
which can be approximated as
\bea\label{P22}
\mathcal{P}_{\mathcal{R}} \sim \frac{\chi^2}{(0.785+\chi)^2},
\eea
where $\chi=\frac{k_0}{a_0 H_0}$. A comparison of the spectrum in Eq.~(\ref{P21}) and ~(\ref{P22}) is plotted in Fig.~(\ref{Fig02}). We can see that the approximated spectrum reproduces the analytical spectrum very well. Thus, the analytical primordial power spectrum can be parameterized as
\bea
\mathcal{P}_{\mathcal{R}}(k)=A_{s}\Big(\frac{k}{k_{*}}\Big)^{n_s-1}\frac{k^3}{k^3_{0}}\frac{\chi^2}{(0.785+\chi)^2}.
\eea
Note that for the flat case $K=0$, we get $k_0=k$ and the result in Ref.~\cite{Labrana2015} is obtained.

\begin{figure*}[htp]
\begin{center}
\includegraphics[width=0.45\textwidth]{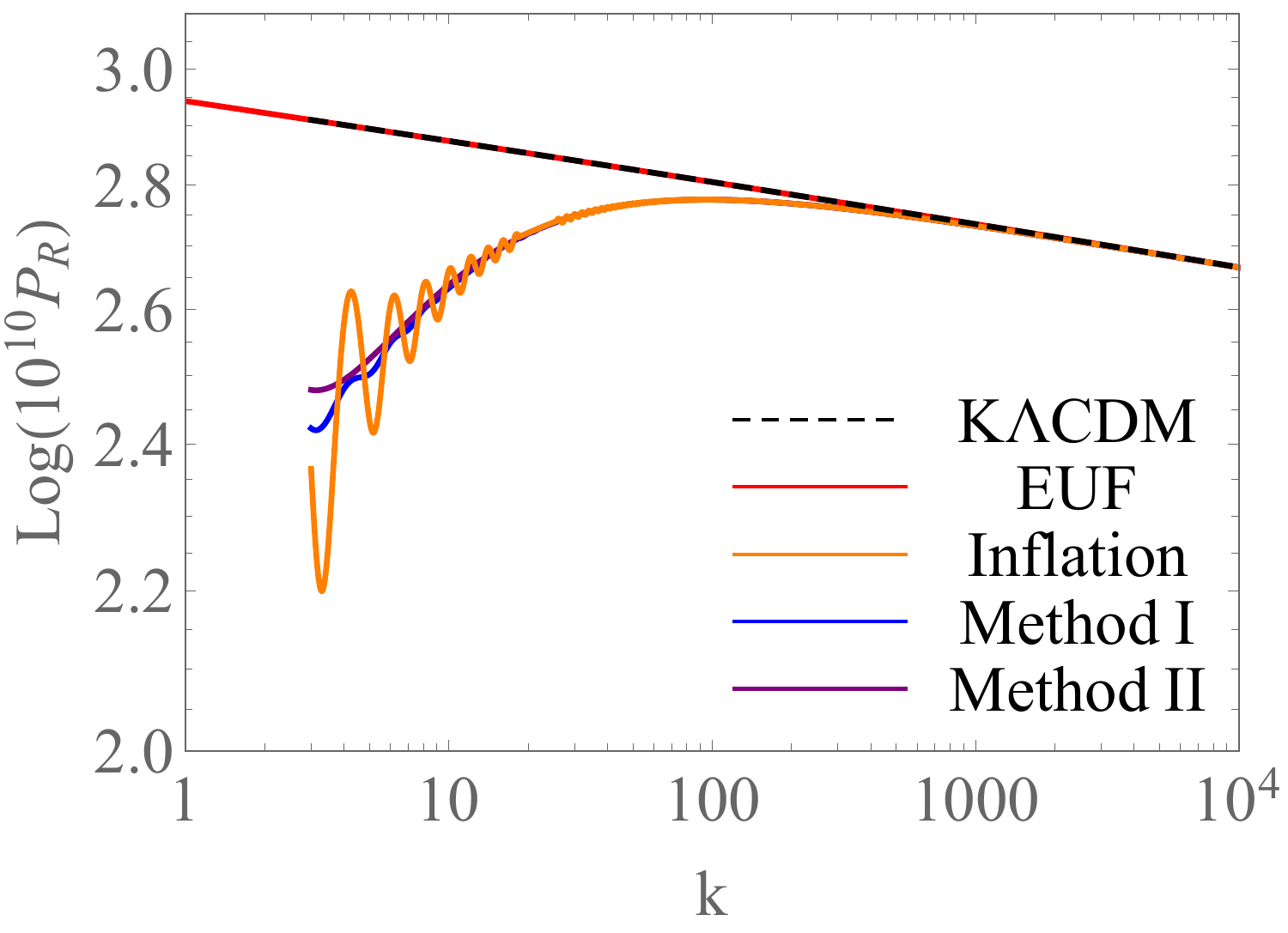}
\includegraphics[width=0.45\textwidth]{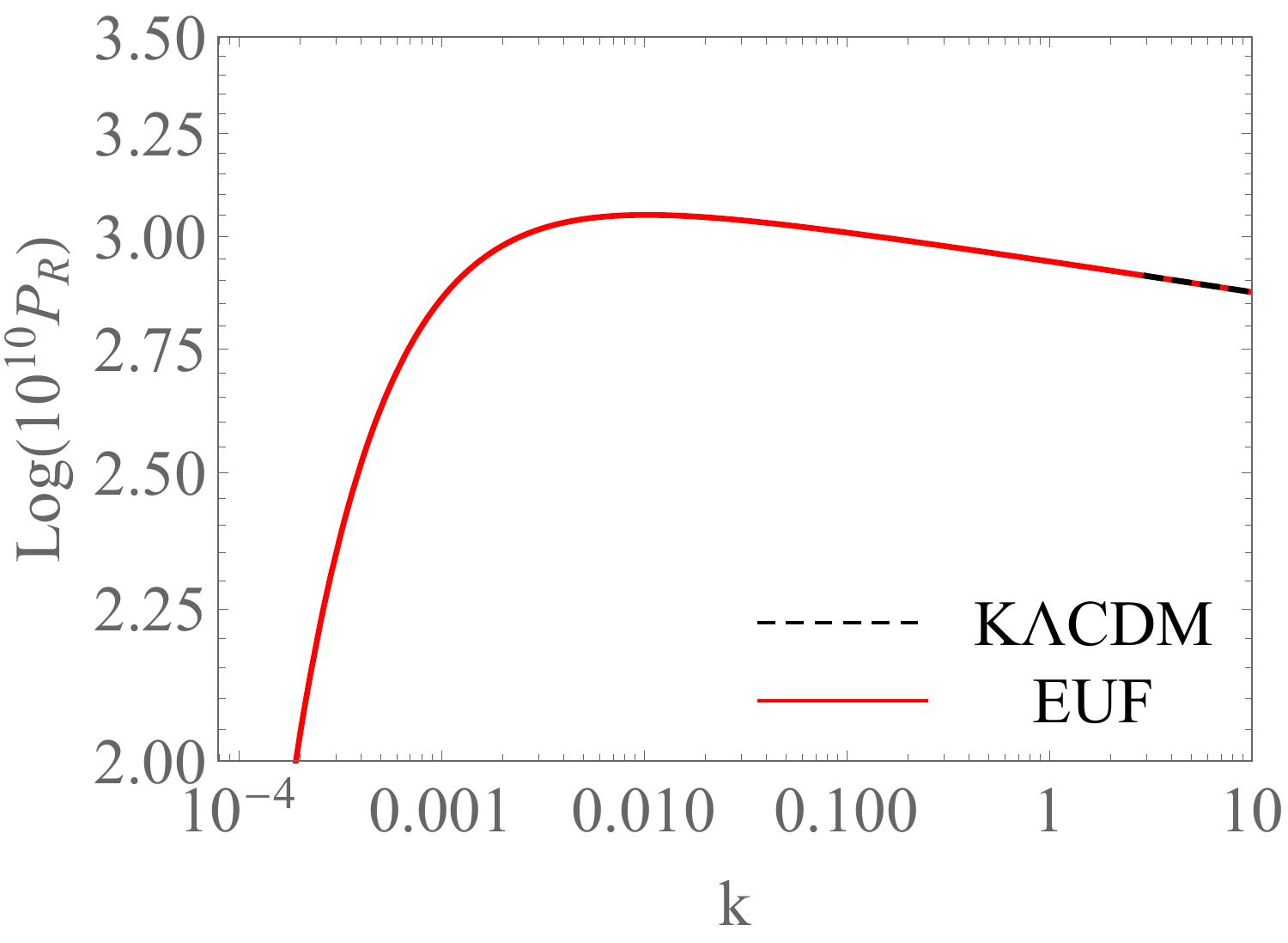}
\caption{\label{Fig03} Primordial power spectra of $\mathcal{P}_{\mathcal{R}}$. The spectrum of EUF is plotted by the red line, the case for K$\Lambda$CDM is depicted by the black dashed line, and the spectrum of inflation shown by the orange line is the case of $\eta_t=\eta_{max}$ in Ref.~\cite{Thavanesan2021}. For the closed universe, only integer values of $k$ with $k\geq3$ are allowed.}
\end{center}
\end{figure*}

The primordial power spectra of $\mathcal{P}_{\mathcal{R}}$ generated by the analytical approximations are shown in Fig.~(\ref{Fig03}). For K$\Lambda$CDM model, the power spectrum is parameterized as
\bea
\mathcal{P}^{K\Lambda CDM}_{\mathcal{R}}(k)=A_{s}\Big(\frac{k}{k_{*}}\Big)^{n_s-1}.
\eea
For the closed universe, we use the Planck 2018 results in the curved universes best-fit data (TT,TE,EE+lowl+lowE+lensing) $A_s=2.0771 \pm 0.1017 \times 10^{-9}$ and $n_s = 0.9699 \pm 0.0090$. For the flat universe, the Planck 2018 results in Ref.~\cite{Planck2020} is used.

In Fig.~(\ref{Fig03}), the spectrum of inflation shown by the orange line corresponds to the case $\eta_t=\eta_{max}$ in Ref.~\cite{Thavanesan2021}, the spectrum of the emergent universe scenario without considering the curvature's contributions on primordial perturbations (EUF) is plotted by the red line, and the spectrum for K$\Lambda$CDM, method I and method II are depicted by the black dashed, blue and purple lines, respectively. From the left panel of Fig.~(\ref{Fig03}), we can see that the spectrum of inflation oscillates for $k<20$ and it has the similar evolutionary behaviors with method I and method II. On the right panel of Fig.~(\ref{Fig03}), the spectrum of EUF is suppressed for $k<10^{-3}$.

In EUF, the suppression of power spectrum is caused by the Einstein static state~\cite{Labrana2015}, while the spatial curvature leads to the suppression of power spectrum in the ultraslow-roll inflation~\cite{Thavanesan2021}. For the spatially closed emergent universe studied in this work, we find that both Einstein static state and the positive spatial curvature can result in the suppression of power spectrum. In the Einstein static state, the Hubble horizon $1/H$ approaches infinity since $H=0$, and all perturbation modes are deeply inside the Hubble horizon. When the universe evolves into the inflationary phase, due to the rapid decreasing of Hubble horizon, the long wavelength modes (low $k$) first leave the Hubble horizon during inflation, and the short wavelength modes (high $k$) exit the Hubble horizon later. As a result, in this paper, the power spectrum for high $k$ is same as that in $\Lambda$CDM model, while it is suppressed for low $k$.

Both panels in Fig.~(\ref{Fig03}) show that a suppression of primordial power spectra exist in low $k$, and this suppression may lead to a suppression on the CMB TT-spectrum. In the following section, we will analyze the CMB TT-spectrum of the emergent universe scenario.

\section{CMB temperature power spectra}

\begin{figure*}[htp]
\begin{center}
\includegraphics[width=0.45\textwidth]{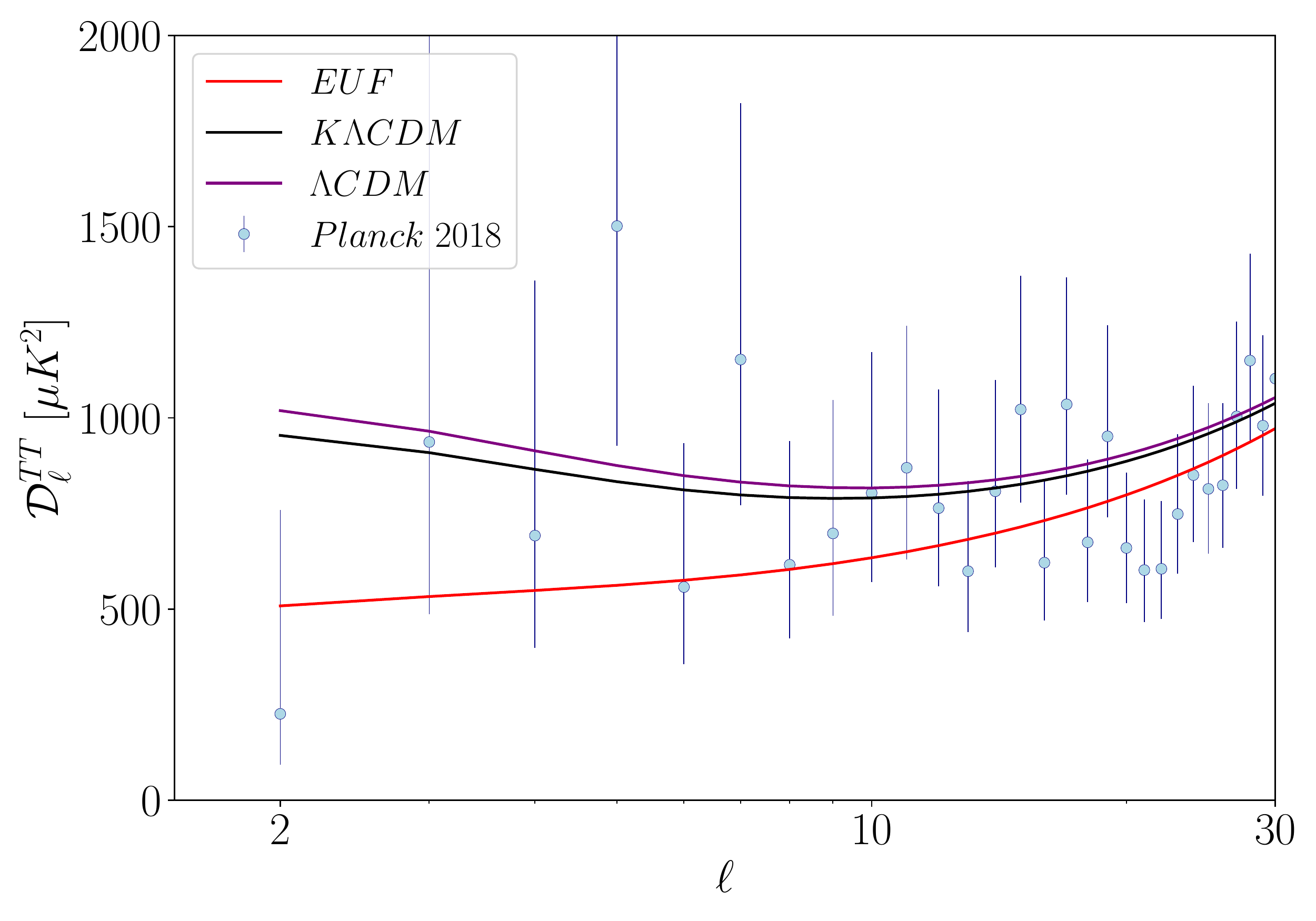}
\includegraphics[width=0.45\textwidth]{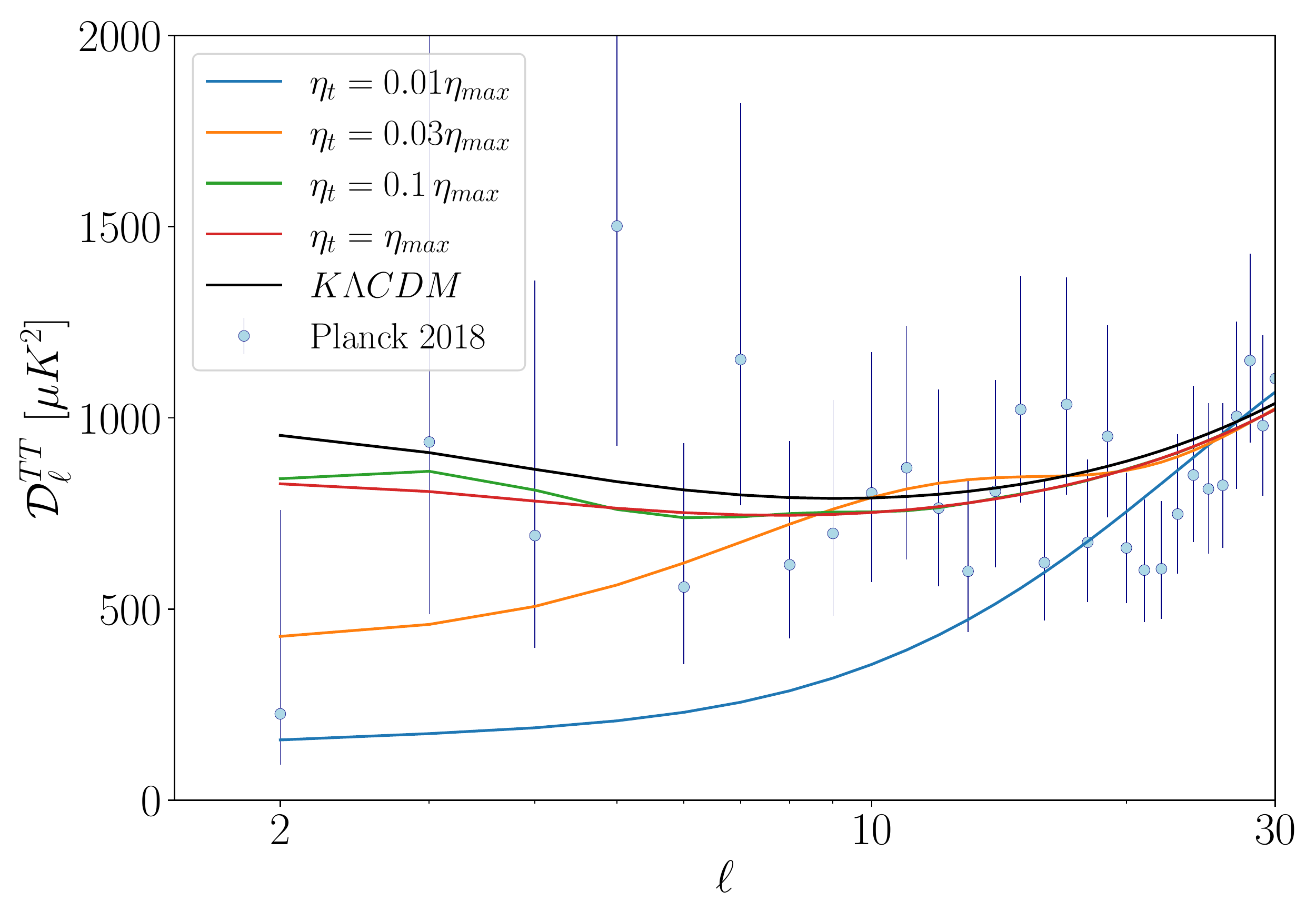}
\includegraphics[width=0.45\textwidth]{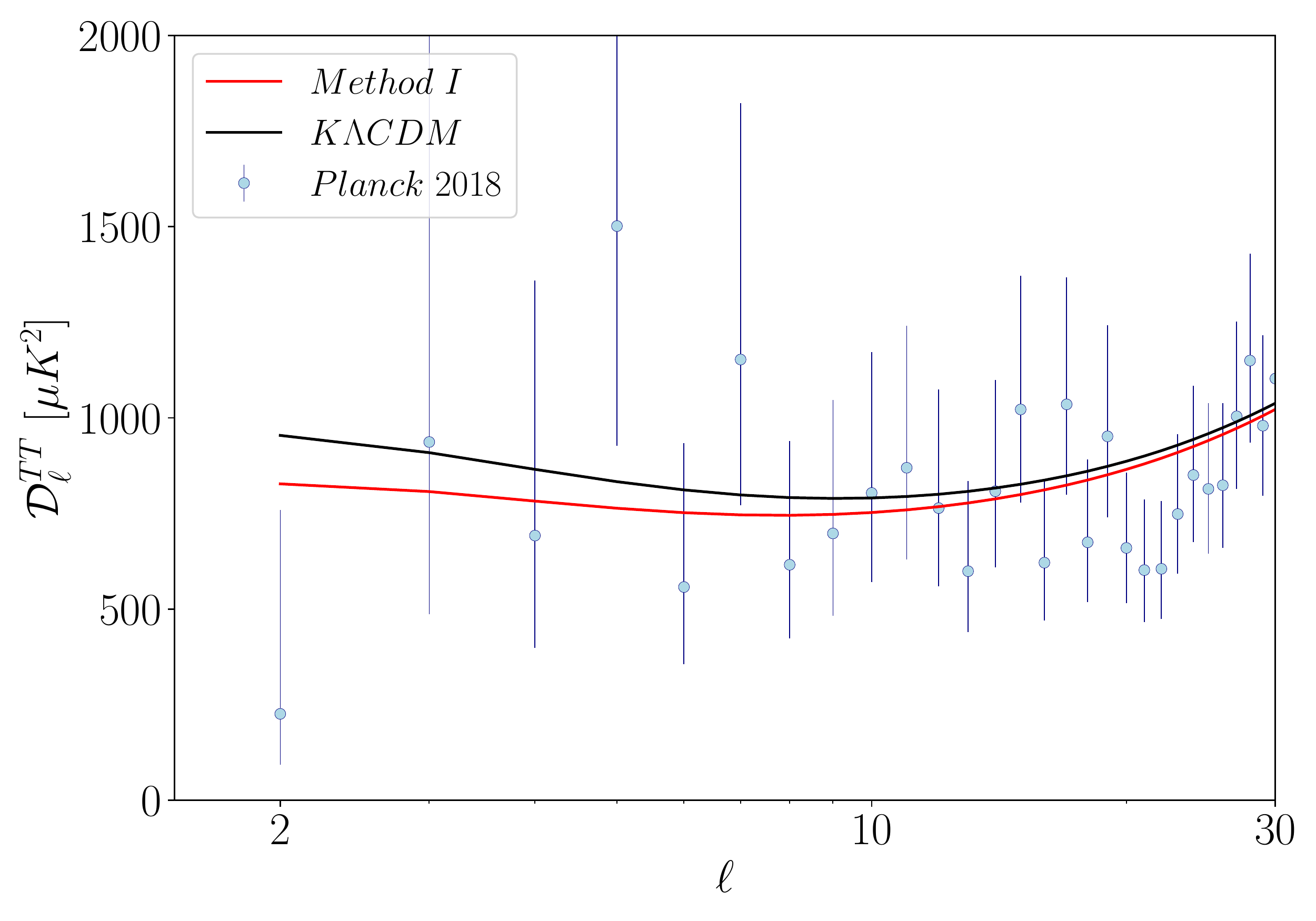}
\includegraphics[width=0.45\textwidth]{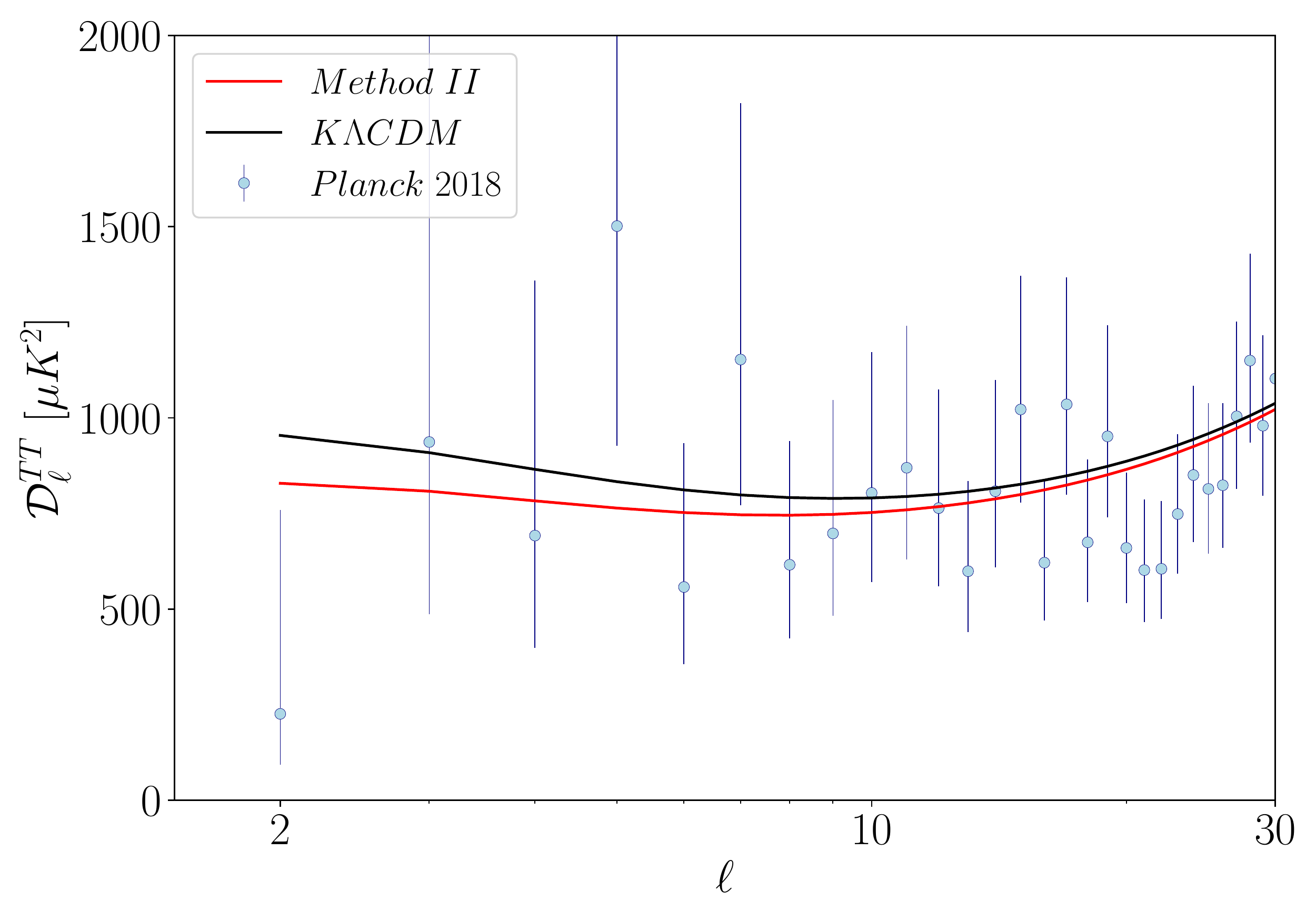}
\caption{\label{Fig04} CMB temperature power spectra for EUF, inflation, method I and method II. The result in the first panel was obtained in Ref.~\cite{Labrana2015} and the second panel are the results of $K=1$ in Ref.~\cite{Thavanesan2021}. The points represent the Planck 2018 data.}
\end{center}
\end{figure*}

\begin{figure*}[htp]
\begin{center}
\includegraphics[width=0.45\textwidth]{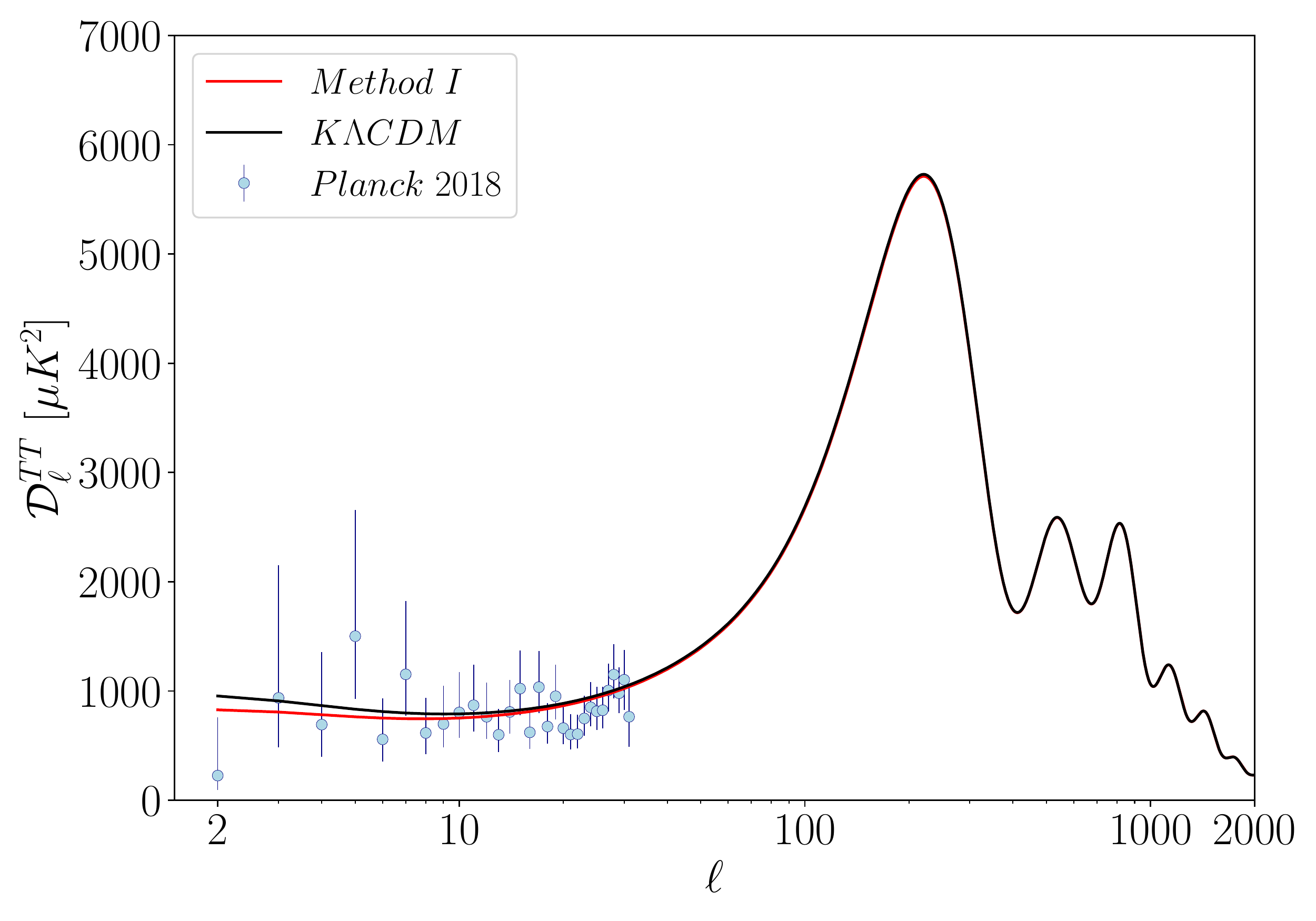}
\includegraphics[width=0.45\textwidth]{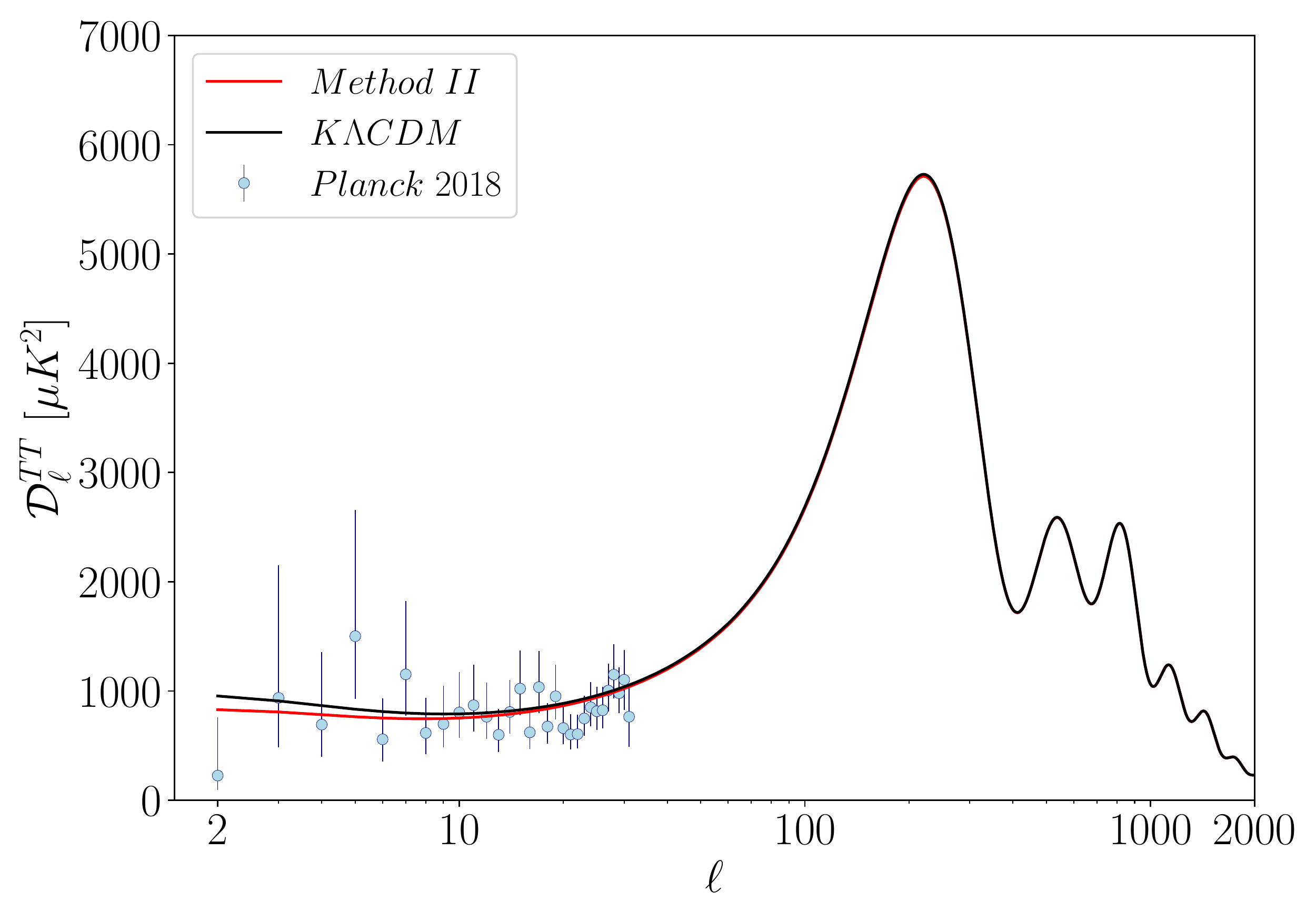}
\caption{\label{Fig05} CMB temperature power spectra for method I and method II. The points represent the Planck 2018 data.}
\end{center}
\end{figure*}

In order to discuss and show the power suppression of the CMB TT-spectrum at large scales coming from the emergent universe, we use the CLASS code~\cite{Blas2011} to plot the CMB TT-spectra of EUF, inflation, method I and method II in Fig.~(\ref{Fig04}). In the first panel, we plot the CMB TT-spectra of EUF, K$\Lambda$CDM and $\Lambda$CDM model. The result in the first panel was obtained by Labrana in Ref.~\cite{Labrana2015} where the space curvature was only responsible for the existence of Einstein static universe and the contributions to the primordial power spectrum were neglected. We can see from the first panel that EUF generates a strong suppression of the CMB TT-spectrum. The CMB TT-spectra in the second panel are the results of $K=1$ in Ref.~\cite{Thavanesan2021} in which the suppression of CMB TT-spectrum is altered by adjusting the transition time $\eta_t$ and a strong suppression occurs for $\eta_t<0.03\eta_{max}$. In the third and forth panels, we depict the CMB TT-spectrum of the method I and method II respectively, and it can be seen that we can not discriminate method I from method II by the CMB TT-spectrum. A more detailed and complete CMB TT-spectrum for method I and method II are shown in Fig.~(\ref{Fig05}). In this figure, we can see that both the spectrum lines of method I and method II are overlapped with the K$\Lambda$CDM for $l>30$, while a suppression of spectrum is visible at $l<30$.

Comparing the CMB TT-spectrum of EUF and method II in Fig.~(\ref{Fig04}), we can see that EUF can generate a strong suppression of the CMB TT-spectrum at large scale while method II can only generate a weak but visible suppression. So, when the effects of spatial curvature on the primordial power spectrum are considered, the strong suppression of CMB TT-spectrum stemming from the Einstein static state is relieved.

Comparing the CMB TT-spectra of method I, method II with the inflation with $\eta_t = \eta_{max}$, we find that the spectra of them are overlapped and we can not distinguish method I or method II from the inflation with $\eta_t = \eta_{max}$. It is worthy to note that the CMB TT-spectrum of inflation is sensitive to the transition time $\eta_t$ and more precise comparison results between method I and inflation requires a definite transition time.

\section{Conclusion}

The emergent universe scenario provides a possible way to solve the big bang singularity by suggesting that the universe originated from an Einstein static state. After the universe exits naturally from the Einstein static state, it evolves into an inflationary epoch. In order to discriminate the emergent universe scenario from the inflation, we need to find out the difference between them from the power spectrum.

In this paper, we study the primordial power spectrum and the CMB TT-spectrum of the emergent universe scenario in the closed FLRW spacetime by two different methods. In method I, the emergent universe scenario consists of two evolutionary phase, i.e. the emergent universe beginning with an Einstein static sate then followed by an instantaneous transition to a ultraslow-roll inflation, while the scale factor obeys a fixed evolutionary way in method II. For method II, the primordial power spectrum depends on the free parameter $a_0 H_0$, and this problem is avoided in method I.

Comparing the primordial power spectra of method I, method II, EUF and the inflation, we find that the spectrum of inflation oscillates at the low $k$, and has the similar evolutionary behavior with method I and method II. The spectra of the four models all suppressed at low $k$: the EUF is suppressed at $k<10^{-3}$ and the other is suppressed at $k<30$.

After analyzing method I and II by the CMB TT-spectrum, we find that the CMB TT-spectra of them are suppressed at $l<30$, and both the spectra of them are nearly identical. Then, by comparing the CMB TT-spectra of method I, method II, EUF and inflation, we find that, method I and method II can not be distinguished from the inflation with the special case $\eta_t = \eta_{max}$ by the CMB TT-spectrum, and the strong suppression of CMB TT-spectrum stemming from the Einstein static state in EUF is relieved when the spatial curvature is considered in method II.

\begin{acknowledgments}

This work was supported by the National Natural Science Foundation of China under Grants Nos. 11865018, 11865019, 11505004, the Foundation of Guizhou Science and Technology Department of China under Grants Nos. QKHJC[2019]1323, the Foundation of the Guizhou Provincial Education Department of China under Grants Nos. KY[2018]028, the Doctoral Foundation of Zunyi Normal University of China under Grants No. BS[2017]07.

\end{acknowledgments}

\end{document}